\documentclass{llncs}

\usepackage{amsmath,amsfonts}
\usepackage{xspace}
\usepackage{txfonts}
\usepackage{proof}

\usepackage{pgf}
\usepackage{tikz}
\usetikzlibrary{arrows,automata}
\usetikzlibrary{positioning}
\usetikzlibrary{shapes}

\usepackage{xspace}

\newcommand{\eg}{\textit{e.g.}}
\newcommand{\etc}{\textit{etc.}}
\newcommand{\ie}{\textit{i.e.}}
\newcommand{\cf}{\textit{cf.}}
\newcommand{\lb}{\texttt{\char'173}} 
\newcommand{\rb}{\texttt{\char'175}} 

\newcommand{\secref}[1]{Sec.~\ref{#1}}
\newcommand{\CTLs}{\ensuremath{\text{CTL}^\ast}\xspace}
\newcommand{\CTLsFO}{\ensuremath{\text{CTL}^\ast(\text{FO})}\xspace}

\newcommand{\calC}{{\cal C}}
\newcommand{\calD}{{\cal D}}
\newcommand{\calF}{{\cal F}}
\newcommand{\calP}{{\cal P}}

\newcommand{\calS}{{\cal S}}
\newcommand{\calI}{{\cal I}}

\newcommand{\Guard}{\mathsf{Guard}}
\newcommand{\GProg}{\mathsf{GProg}}
\newcommand{\Update}{\mathsf{Update}}
\newcommand{\db}{\ensuremath{\mathit{db}}\xspace}
\newcommand{\DB}{\ensuremath{\mathit{DB}}\xspace}
\newcommand{\sfdb}{\ensuremath{\mathsf{db}}\xspace}

\newcommand{\entry}{\ensuremath{\mathrm{entry}}\xspace}
\newcommand{\exit}{\ensuremath{\mathrm{exit}}\xspace}
\newcommand{\init}{\ensuremath{\mathrm{init}}\xspace}

\newcommand{\pqE}{\mathsf{E}}
\newcommand{\pqA}{\mathsf{A}}

\newcommand{\toU}{\mathsf{U}}
\newcommand{\toR}{\mathsf{R}}
\newcommand{\toX}{\mathsf{X}}
\newcommand{\toF}{\mathsf{F}}
\newcommand{\toG}{\mathsf{G}}
\newcommand{\toWX}{\mathsf{\overline{X}}}
\newcommand{\toWU}{\mathsf{W}}

\newcommand{\ltlU}{\toU}
\newcommand{\ltlX}{\toX}
\newcommand{\ltlF}{\toF}
\newcommand{\ltlG}{\toG}
\newcommand{\ltlWX}{\toWX}
\newcommand{\ltlWU}{\toWU}
\newcommand{\ltlR}{\toR}

\def\void{}
\newcommand{\infrule}[3][\void]{%
  {\renewcommand\arraystretch{1.25}
    \ifx\void#1\else\IR{#1}\hspace{0.5em}\fi
    \begin{array}[c]{@{\hspace*{1em}}c@{\hspace*{1em}}}#2\\\hline #3
    \end{array}}}

\newcommand{\IR}[1]{\text{\textsf{#1}}\xspace}

\ifnum\pdfoutput>0
\else
\usepackage[active]{srcltx} 
\fi

\begin{document}
\setlength\abovedisplayskip{5pt plus 2pt minus 4pt}
\setlength\belowdisplayskip{5pt plus 2pt minus 4pt}

\title{Reasoning with Data-Centric Business Processes}

\author{Andreas Bauer \and Peter Baumgartner \and Michael Norrish}
\institute{NICTA%
  \thanks{\scriptsize NICTA is funded by the Australian Government as
    represented by the Department of Broadband, Communications and the
     Digital Economy and the Australian Research Council through the ICT
     Centre of Excellence program.} Software Systems Research Group,
  and\\The Australian National University}

\maketitle

\pagestyle{plain}


\begin{abstract}
We describe an approach to modelling and reasoning about data-centric business processes and present a form of general model checking.
Our technique extends existing approaches, which explore systems only from concrete initial states.

Specifically, we model business processes in terms of smaller fragments, whose possible interactions are constrained by first-order logic formulae.
In turn, process fragments are connected graphs annotated with instructions to modify data.
Correctness properties concerning the evolution of data with respect to processes can be stated in a first-order branching-time logic over built-in theories, such as linear integer arithmetic, records and arrays.

Solving general model checking problems over this logic is considerably harder than model checking when a concrete initial state is given.
To this end, we present a tableau procedure that reduces these model checking problems to first-order logic over arithmetic. The resulting proof obligations are passed on to appropriate ``off-the-shelf'' theorem provers.
We also detail our modelling approach, describe the reasoning components and report on first experiments.

\end{abstract}


\section{Introduction}
Data is becoming increasingly important to large organisations, both private enterprises and large government departments.
Recent headlines on ``big data'' (\cf~\cite{NYT120212}) suggest that many organisations manage unprecedented amounts of structured data, and that worldwide, the volume of information processed by machines and humans doubles approximately every two years.
Organisations need to be able to organise and process data according to their defined business processes, and according to business rules that may further specify properties of the processed data.

Unfortunately, most approaches to business process modelling do not adequately support the analysis of the complex interactions and dependencies that exist between an organisation's processes and data.
Although they may support process analysis, helping users find and remove errors in their models, most fall short when the processes are closely tied to structured data.
The reasons for this are specific to the concrete formalism used for the analyses, but can normally be traced back to the fact that classical propositional logic or discrete Petri-nets are used.
Neither of these can adequately represent structured data and the operations on it.
In other words, these tools' analyses make coarse abstractions of the data, and instead focus mostly on the correctness of workflows.

The business artifact approach, initially outlined in \cite{DBLP:journals/ibmsj/NigamC03}, was one of the first to tackle this issue.
It systematically elevates data to be a ``first-class citizen'', while still offering automated support for process analysis.
Its cornerstones are
\emph{artifacts}, which are records of data values that can change over
time due to the modifications performed by \emph{services}, which are
formalised using first-order logic.
Process analysis is provided, essentially, by means of model checking.
That is, the following question is answered automatically: given some artifact model, a database providing initial values, and a correctness property in terms of a first-order linear-time temporal logic formula (called LTL-FO), do all possible artifact changes over time satisfy the correctness property?
For the constraints given in~\cite{DBLP:conf/bpm/DamaggioDHV11}, this problem is always decidable.

In this paper, we present an approach to modelling and reasoning about data-centric business processes, which is similar to this work, but which offers reasoning support that goes beyond that work's ``concrete model checking''.
Our approach is based on \emph{process fragments} that describe specific tasks of a larger process, as well as \emph{constraints} for limiting the interactions between the fragments.
As such it is also inspired by what is known as \emph{declarative
  business process modelling}~\cite{DBLP:conf/bpm/PesicA06}, meaning
that users do not have to create a single, large transition system
containing all possible task interleavings.
Instead, users can create many small process fragments whose interconnections are governed by rules that determine which executions are permitted.

In our framework, those rules are given by first-order temporal logic.
Unlike~\cite{DBLP:conf/bpm/DamaggioDHV11}, we choose to extend \CTLs, \ie, a branching time logic, rather than LTL, since process fragments are essentially annotated graphs and \CTLs is, arguably, an appropriate formalism to express its properties (\cf~\cite{clarke_em-etal:1999a}).
Our database is given in terms of JSON objects~\cite{JSON}, enriched by a custom, static type system which models and preserves the type information of any input data.
Process fragments may modify data, and one can easily state and answer the concrete model checking problem as outlined above.

However, our approach also works if
one does not start with an initial concrete database; that is, we intend to not only check
whether it is possible to, reach a bad state (\eg, a set of data for
which no process fragment is applicable) from some given state (\ie,
the initial set of data), but also to determine whether for \emph{any}
set of data a bad state can be reached.
In other words, we support what we call \emph{generic model checking}.
As the domains of many data items are infinite (\eg, any item of type integer), this problem is considerably harder, in fact, generally undecidable.

Informally, the two reasoning problems we are interested in are:
\begin{description}
\item[Concrete data model checking problem:] Given a specification $\calS$, a
  database $s_0$, and a \CTLsFO  formula $\Phi$. Does
  $(s_0,\calS) \models \Phi$ hold?
\item[Unrestricted model checking problem:] Given a specification $\calS$ and
  a \CTLsFO  formula $\Phi$. For every database $s_0$, does $(s_0,\calS) \models \Phi$ hold?
\end{description}
As will become clear below, a \emph{specification} is comprised of a process model,
logical definitions, and constraints to combine process fragments. The relation 
$(s_0,\calS) \models \Phi$ means that the pair $(s_0,\calS)$ satisfies the query $\Phi$.
See Section~\ref{sec:process} for the precise semantics.

Without any further restrictions, both problems are not even semi-decidable.  This
can be seen, \eg, by reduction from the domain-emptyness problem of 2-register
machines. Hence, practical approaches need to work with restrictions to recover more
pleasant complexity properties.


The rest of the paper is structured as follows.
In Section~\ref{sec:example} we present a running example.
In Section~\ref{sec:data} we explain the way we handle the rich data of our models: with JSON values, a special type system for those values, and a sorted first order logic for further constraining and describing those values.
This much covers business \emph{rules}; in Section~\ref{sec:process}, we describe how we can model \emph{processes}.
When processes (actually process \emph{fragments}) combine with rules, we get what we call \emph{specifications}.
In Section~\ref{sec:ctlsfo-tableaux}, we describe the tableau-based model checking algorithm that is used to decide user queries of the two sorts identified above.
Section~\ref{sec:experiments} discusses how we have implemented our technology, and describes some experimental results.
Finally, we conclude in Section~\ref{sec:conclusion}.



\section{A Running Example: Purchase Order}
\label{sec:example}

\begin{figure}[tbp]
\textbf{Process model:}\\
\begin{tikzpicture}
  [
   node distance=1.8cm,
   state/.style={
     text width=1.7cm, rectangle, rounded corners, draw=black,
     very thick, minimum height=2em, inner sep=2pt, text centered},
   task/.style={
     rectangle, rounded corners, shade, top color=white,
     bottom color=blue!50!black!20, draw=blue!40!black!60,
     very thick}
  ]


 \node[state,task]                      (INIT)      { Init };
 \node[state,task,below left of=INIT]   (PACK)      { Pack };
 \node[state,task,below of=PACK]        (STOCKTAKE) { Stocktake };
 \node[state,task,below right of=INIT]  (DECLINED)  { Declined };
 \node[state,task,below of=DECLINED]    (PACKED)    { Packed };
 \node[state,task,below of=PACKED]      (INVOICED)  { Invoice };

 \path[->] (INIT) edge                    node[left]  {$e_1$} (PACK);
 \path[->] (INIT) edge                    node[right](DECLINED_EDGE) {$e_2$} (DECLINED);
 \path[->] (PACK) edge[bend left=25]      node[right] {$e_4$} (STOCKTAKE);
 \path[->] (STOCKTAKE) edge[bend left=25] node[left]  {$e_3$} (PACK);
 \path[->] (PACK) edge                    node[right] {$e_5$} (PACKED);
 \path[->] (PACKED) edge                  node[right] {$e_6$} (INVOICED);


 \node[state, task, right of=INIT, yshift=-2.8cm,xshift=2.2cm] (PAID)      { Paid };
 \node[below of=PAID,yshift=0.5cm]                             (PAID_INCOMING) { };
 \path[->] (PAID_INCOMING) edge node[right](PAID_EDGE) {$e_7$} (PAID);

 \node[state, task, right of=PAID,xshift=0.8cm]        (SHIPPED)   { Shipped };
 \node[below of=SHIPPED,yshift=0.5cm]                  (SHIPPED_INCOMING) { };
 \path[->] (SHIPPED_INCOMING) edge node[right] {$e_8$} (SHIPPED);

 \node[state, task, right of=SHIPPED,xshift=0.8cm]       (COMPLETED) { Completed };
 \node[below of=COMPLETED,yshift=0.5cm]                  (COMPLETED_INCOMING) { };
 \path[->] (COMPLETED_INCOMING) edge node[right] {$e_9$} (COMPLETED);


 \node[above of=PAID,yshift=-0.4cm,xshift=1.1cm,rectangle split parts=5,
       draw,text width=2.5cm] (LEGEND_PAID)
       {
         entry = ``$true$''\\
         exit = ``$true$''
       };
 \draw[dashed] (LEGEND_PAID) -- (PAID);
 \draw[dashed] (LEGEND_PAID) -- (SHIPPED);

 \node[below of=PAID,yshift=-0.25cm,xshift=2.5cm,rectangle split parts=5,
       draw,text width=5.3cm] (LEGEND_PAID_EDGE)
       {
         guard = ``\texttt{db.status.paid <> true}''\\
         script = ``\texttt{db.status.paid = true}''
       };
 \draw[dashed] (LEGEND_PAID_EDGE) -- (PAID_EDGE);

 \node[above of=COMPLETED,yshift=-0.4cm,xshift=-0.6cm,rectangle split parts=5,
       draw,text width=2.5cm] (LEGEND_COMPLETED)
       {
         entry = ``$true$''\\
         final = ``$true$''
       };
 \draw[dashed] (LEGEND_COMPLETED) -- (COMPLETED);

 \node[right of=INIT,yshift=0cm,xshift=3.5cm,rectangle split parts=3,
       draw,text width=5.2cm] (LEGEND_DECLINED)
       {
         guard = ``\texttt{$\sim$acceptable(db)}'' \\
         script = ``\texttt{db.status.final = true}''
       };
 \draw[dashed] (LEGEND_DECLINED) -- (DECLINED_EDGE);
\end{tikzpicture}

\textbf{Definitions:}

completed: $\forall \mathtt{s}\text{:}\mathtt{Status}\ .\ (\mathtt{completed}(\mathtt{s}) \Leftrightarrow ( \mathtt{s}.\mathtt{paid} = \mathtt{true} \wedge \mathtt{s}.\mathtt{shipped} = \mathtt{true} ))$

accepted: $\forall \mathtt{db}\text{:}\mathtt{DB}\ .\ (\mathtt{acceptable}(\mathtt{db}) \Leftrightarrow (\neg \mathtt{isEmpty}(\mathtt{db}.\mathtt{order})))$

readyToShip: $\forall \mathtt{s}\text{:}\mathtt{Status}\ .\ (\mathtt{readyToShip}(\mathtt{s}) \Leftrightarrow ( \mathtt{isEmpty}(\mathtt{s}.\mathtt{open})))$ \ldots\\

\textbf{Constraints:}

nongold: $(\mathtt{db}.\mathtt{gold} = \mathtt{false} \Rightarrow (\mathtt{db}.\mathtt{status}.\mathtt{shipped} = \mathtt{false}\, \toWU\, \mathtt{db}.\mathtt{status}.\mathtt{paid} = \mathtt{true}))$

\caption{Model of a purchase order system as process fragments and definitions.}
\label{fig:purchase}
\end{figure}

In this section, we introduce a simplified model of a purchase order
system using process fragments.  The purpose of the modelled system is
to accept incoming purchase orders and process them further (packing,
shipping, etc.), or to decline them straight away if there are
problems.
The whole model is depicted as a graph in Fig.~\ref{fig:purchase}, where
the biggest process fragment is on the left, with further atomic
fragments beside it (labelled Paid, Shipped, and Completed,
respectively).
Both process tasks, represented as nodes in the graph,
and connections are typically annotated with extra information.
Node annotations determine whether or not a node is an initial and/or a final node, an entry and/or an exit node.
This information is used to constrain the ways in fragments can connect.
Edges can carry a guard given as a
formula and a simple program written in the programming language Groovy.
The purpose of the program (given in the field ``script'') is to modify
the underlying database, which is referred to by the variable \verb+db+.

The depicted system model has one initial node, Init, where it
waits for a purchase order to arrive.
Then, the system can either start to pack (\ie, enter node Pack), or decline the order (\ie, enter node
Declined).
An order can be declined if the guard ($\neg \mathit{acceptable}(\db)$) in the annotation of edge $e_2$ is satisfied.
The predicate $\mathit{acceptable}$ is defined in the
Definitions section of our input specification.
In a nutshell, the sections Definitions and Constraints contain domain-knowledge, encoded as logical rules.
(The constraint named ``nongold'' states that non-gold customers
must pay before shipment; $\toWU$ is the ``weak until'' operator.)

If the order is not declined, an attempt will be made to pack its constituents.
If all are in stock, the process will continue to the node Packed.
However, if one or more items are missing, they need to be ordered in, which is expressed in the loop between the nodes Pack and Stocktake.

Informally, process fragments are linked together as follows.
Starting from a state comprised of an \init node and a given initial database, an outgoing transition from the current state can only be executed
if it satisfies the transition's guard.
If it is satisfied, the associated program is executed to determine the new value of the database, and the edge's target node becomes the new current state.
The \entry and \exit annotations impose implicit constraints on how
fragments can be combined: the execution of a new process fragment must
always start with its \entry node\footnote{For simplicity we assume
  every fragment contains exactly one entry node.} coming from an \exit node. In
other words, there are implicit transitions between all \exit and all \entry nodes. 
However, if a guard is associated to an
\entry \emph{node}, this guard sits on all its implicit incoming transitions. 
The computation stops if from the current
state no successor can be reached, either because there is no outgoing
edge, the guards of all outgoing edges are not satisfied by the
current state, or a depth limit has been reached.

In our example, two possible sequences are Init $\rightarrow$ Declined, or
Init $\rightarrow$ Pack $\rightarrow$ \ldots $\rightarrow$ Invoice
$\rightarrow$ Paid $\rightarrow$ Shipped$ \rightarrow$ Completed.
It is not required to cover all fragments, as illustrated by the first run.

The database which can be modified by the programs given in the
``script'' annotations, is represented as a JSON object.  See, for
example, left hand side of Fig.~\ref{fig:data}.  (The right hand side
contains type definitions for the JSON data, see also
\secref{sec:data}.)  The program annotated on
edge $e_2$, which leads into node Declined, simply sets the field
\texttt{final} inside \texttt{status} to \texttt{true}.
Crucial for our example is the list of open items, under
\texttt{status}, which has to be empty to be able to ship a purchase
order.  If it is not, constituents of the order are missing and need to
be ordered until the list is empty.

As for sample queries consider the \CTLsFO formula
$ \neg (\pqE\, \toF\, \db.\mathit{status}.\mathit{final}=\mathit{true})$, which can be seen as a \emph{planning} goal.
The runs on the model above that \emph{falsify} it lead to a database $\db$ that has reached a ``final'' state, with $\mathit{status}.\mathit{final}$ being set to $\mathit{true}$.
Planning queries are useful, \eg, for flexible
process configuration from fragments during runtime. Another interesting query is
$\pqA\, \toG\, (\forall  s\text{:}Stock\, . \, (s \in  db.stock \Rightarrow  s.available \ge 0))$.
It is a safety property, saying that at all stages in the process run, and for all possible stock items, the number of available items is non-negative.
Such queries are typical during design time, and pose an unrestricted model checking problem.

\begin{figure}[tbp]
\begin{minipage}[t]{0.5\linewidth}
\begingroup
\fontsize{7pt}{9pt}\selectfont
\begin{verbatim}
{   "order" : [1],
    "gold"  : true,
    "stock" : [ { "ident" : "Mouse",
                  "price" : 10,
                  "available" : 0  },
                { "ident" : "Monitor",
                  "price" : 200,
                  "available" : 2  },
                { "ident" : "Computer",
                  "price" : 1000,
                  "available" : 4  } ],
    "status" : {  "open" : [],
                  "value" : 0,
                  "shipping" : 0,
                  "paid" : false,
                  "shipped" : false,
                  "final" : false }    }
\end{verbatim}
\endgroup
\end{minipage}
\begin{minipage}[t]{0.5\linewidth}
\begingroup
\fontsize{7pt}{9pt}\selectfont
\begin{verbatim}
   DB  = { order: List[Integer],
           gold: Bool,
           stock: List[Stock],
           status: Status      }

 Stock = { ident: String,
           price: Integer,
           available: Integer  }

Status = { open: List[Integer],
           value: Integer,
           shipping: Integer,
           paid: Bool,
           shipped: Bool,
           final: Bool         }
\end{verbatim}
\endgroup
\end{minipage}
\caption{\emph{Left:} Example database as JSON document. \emph{Right:} JSON type constraints.}
\label{fig:data}
\end{figure}


\section{Modelling Data With JSON Logic}
\label{sec:data}
Faithful modelling of business processes requires being able to model the objects (or \emph{data}) manipulated by the processes and, of course, their evolution over time.
In this section we focus on data modelling, which is based on JSON extended with a type system.

JSON~\cite{JSON} is simple, standardised, textual data representation format.  In
addition to a standard set of atomic values such as integers and strings, JSON
supports two structuring techniques: sequencing (``arrays'') and arbitrarily nested
hierarchies (through ``objects'').
Our choice of JSON (rather than XML, say), is based on the ease with which it can be
written and understood by humans.  JSON is sufficiently rich to be a plausible format
for representing the data used in business processes, and its human ease-of-use is
extremely helpful. 

Other than simply being the medium in which data is represented, there are two
important functions that JSON must support.  Firstly, it must be possible to
manipulate JSON values in the course of executing a specification. This
functionality is realised through the use of the Groovy programming notation.

Secondly, it must be possible to express \emph{logical} predicates over JSON values,
both to guard process transitions and to pick out certain forms of value that are of
interest.  In particular, if a specification is to achieve a particular end-goal,
with a database being in a particular configuration, we need to be able to describe
how the various values in that database inter-relate.  It is this that motivates our
choice of the logically expressive capabilities of first order logic, together with
sorts such as lists and numbers.

In addition to first-order predicates, we also use a simple type system over JSON
values.  This provides a simple mapping into the sorts of our underlying first-order logic.
We note that the type system is indispensable for unrestricted model checking, in
order to derive from it logical axioms for object and array manipulations.

\subsection{A Type System for JSON}
First we briefly summarise the syntax that is fully described in the IETF~RFC~\cite{JSON}:
JSON values can be numbers, booleans (\texttt{true} and \texttt{false}), strings (written between double-quotes, \eg, \texttt{"a~string"}) and a special value \texttt{null}.
JSON's \emph{arrays} are written as comma-separated values between square-brackets, \eg, \texttt{[1,~"string",~[true]]}.
JSON \emph{objects} are similar to records or structures in languages such as Pascal and C.
They are written as lists of field-name/value pairs between braces.
Both forms are illustrated in Figure~\ref{fig:data}.

\newcommand{\obj}[1]{\texttt{Obj}\lb#1\rb}
Sibling field-names within an object should be unique, and are considered unordered.
Therefore, an object can be thought of as a finite map from field-names to further JSON values.
Following this conception, we write \obj{$\mathit{vf}$} to denote an object whose field names are the domain of finite map $\mathit{vf}$, with field $s$'s value being $\mathit{vf}(s)$.

\newcommand{\intty}{\texttt{Integer}}
\newcommand{\boolty}{\texttt{Bool}}
\newcommand{\stringty}{\texttt{String}}
\newcommand{\listtyopname}{\texttt{List}}
\newcommand{\listty}[1]{\listtyopname[#1]}
\newcommand{\optiontyopname}{\texttt{Option}}
\newcommand{\optionty}[1]{\optiontyopname[#1]}
\newcommand{\objtyopname}{\texttt{ObjTy}}
\newcommand{\objty}[1]{\objtyopname\lb#1\rb}
\newcommand{\enumtyopname}{\texttt{EnumTy}}
\newcommand{\enumty}[1]{\enumtyopname[#1]}
\newcommand{\dom}{\mathrm{dom}}

JSON does not impose any restrictions on the structure of values.  For example, a
list may contain both strings and integers.  However, we choose to restrict this
freedom with a simple type system comparable to those in third-generation languages
such as C.  Let JSON types be denoted by $\tau$, $\tau'$, $\tau_1$ \etc, then
{\small
\[
\begin{array}{rcl}
\tau &\quad::=\quad & \intty \;\;|\;\; \boolty \;\;|\;\; \stringty \;\;|\;\; \listty{\tau} \;\;|\;\; \optionty{\tau} \;\;|\;\; \objty{\mathit{tf}} \;\;|\;\;  \enumty{\mathit{sl}}
\end{array}
\]
}
where $\mathit{tf}$ is a finite map from strings to types, and $\mathit{sl}$ is a list of strings.

The \optiontyopname{} and \enumtyopname{} types are the only ones that do not have a obvious connection back to a set of JSON values.
The \optiontyopname{} type is used to allow for values that are not necessarily always initialised, but which come to acquire values as a process progresses.
We do not expect to see the option-constructor occur with multiple nestings, \eg, a type such as $\optionty{\optionty{\stringty}}$.
The \enumtyopname{} type is used to model finite enumerated types, where each value is represented by one of the strings in the provided list.
This flexibility in the type system allows for more natural modeling.

Values are assigned types with the following inductive relation, where we write $v :
\tau$ to indicate that JSON value $v$ has type $\tau$, where the meta-variables $i$ and $s$
correspond to all possible integer and string values respectively, and where we use
$e \in \ell$ to mean that element $e$ is a member of list $\ell$:
{\small
\[
\begin{array}{c}
\infer{\texttt{true} : \boolty}{} \qquad \infer{\texttt{false} : \boolty}{} \qquad \infer{i : \intty}{}  \qquad \infer{s : \stringty}{}
\\[3mm]
\infer{s : \enumty{\mathit{sl}}}{s \in \mathit{sl}} \qquad
\infer{\texttt{null} : \optionty{\tau}}{} \qquad
\infer{v : \optionty{\tau}}{v : \tau}
\\[3mm]
\infer{[ \mathit{els} ] : \listty{\tau}}{
  \forall v \in \mathit{els}. \;v : \tau}
\\[3mm]
\infer{\obj{\mathit{vf}} : \objty{\mathit{tf}}}{
  \dom(\mathit{vf}) = \dom(\mathit{tf}) &
  \forall s \in \dom(\mathit{vf}). \;\mathit{vf}(s) : \mathit{tf}(s)
}
\end{array}
\]}%
This type system is simple and designed to be pragmatic.  Meta-theoretically, it is
not particularly elegant.  In particular, values may have multiple types: if a value
$v$ is of type $\tau$, then it is also of type $\optionty{\tau}$; string values are not
just of type $\stringty$, but also have an arbitrary number of possible enumeration
types.

\subsection{From JSON to First-Order Logic}
When a user develops a business specification, we expect them to name the various types of
interest with the type system above. When concrete
initial values are given for a concrete model-checking problem, we use that type
system to check that these values really do have the appropriate type.  The
same system is used to ensure that logical guards and goal-conditions are sensible,
as discussed below. It also plays a pivotal role in our reasoning procedure for the
unrestricted model checking problem, which requires to reflect the semantics of a
JSON type model in many-sorted first-order logic. We are going to describe that now.

We fix a non-empty set $S$ of \emph{sorts} and a first-order logic signature $\Sigma$
comprised of function and predicate symbols of given arities over $S$. We assume
infinite supplies of variables, one for every sort in $S$.
A \emph{constant} is a 0-ary function symbol. The (well-sorted $\Sigma$-)terms and atoms are defined as
usual. We assume $\Sigma$ contains a predicate symbol $\approx_s$ (equality) of arity $s \times
s$, for every sort $s \in S$. Equational atoms, or just \emph{equations}, are written
infix, usually without the subscript
$s$, as in $1+1\approx 2$.  We write $\phi[x]$ to indicate that every free variable in the
formula $\phi$ is among the list $x$ of variables, and we write $\phi[t]$ for the formula
obtained from $\phi[x]$ by replacing all its free variables $x$ by the corresponding
terms in the list $t$.

We assume a sufficiently rich set of Boolean connectives (such as $\{\neg{}, {}\land{} \}$)
and the quantifiers $\forall$ and $\exists$. The \emph{well-sorted
  $\Sigma$-formulas}, or just \emph{(FO) formulas} are defined as usual.  We are
particularly interested in signatures containing (linear) integer arithmetic. For
that, we reserve the sort symbol $\mathbb{Z}$, the constants $0, \pm 1, \pm 2, \ldots$, the
function symbols $+$ and $-$, and the predicate symbol $>$, each of the expected
arity over $\mathbb{Z}$.

The semantics of our logic is the usual one: a \emph{$\Sigma$-interpretation} $I$
consists of non-empty, disjoint sets, called \emph{domains}, one for each sort in
$S$.  We require that the domain for $\mathbb{Z}$ is the set of integers, and
that every arithmetic function and predicate symbol is mapped to its obvious function
over the integers.  A \emph{(variable) assignment $\alpha$} is a mapping from the
variables into their corresponding domains. Given a formula $\Phi$ and a pair $(I,\alpha)$ we
say that \emph{$(I,\alpha)$ satisfies $\Phi$}, and write $(I,\alpha) \models \Phi$, iff
$\Phi$ evaluates to true under $I$ and $\alpha$ in the usual sense (the component $\alpha$ is
needed to evaluate the free variables in $\Phi$). If $\Phi$ is closed then $\alpha$ is
irrelevant and we can write $I \models \Phi$ instead of $(I,\alpha) \models \Phi$. We say that a closed
sentence $\Phi$ is \emph{valid} (\emph{satisfiable}) iff $I \models \Phi$ for all (some)
interpretations $I$. 

In order to map our JSON modelling framework to FOL we let the sorts $S$ contain all
the defined type names in the JSON type model of the given specification. In the
example in Section~\ref{sec:example} these are $\mathtt{DB}$, $\mathtt{Stock}$ and
$\mathtt{Status}$. Without loss of generality we assume that the top-level type in a
JSON type model is always called $\mathtt{DB}$.\footnote{We need additional sorts,
  \eg, for truth values and integers, as mentioned. The sorts in $S$ are written in
  italics, as in $\DB$.}  We call any JSON term of type $\mathtt{DB}$ a
\emph{database}. See again Section~\ref{sec:example} for an example.
We fix a dedicated variable $\db$ of sort $\DB$. Informally, $\db$ will be used to hold
the database at the current time point.

Furthermore, we must provide mappings into FOL from terms that are specific to JSON.
In some sense,
both JSON's arrays and its objects are generic ``arrays'', values that can be seen as
collections of independently addressable components.
The JSON syntax for that is a usual one:  \texttt{a[i]},
denotes the value of the \verb+i+$^{\mathrm{th}}$ element of array \verb+a+; and
\texttt{obj.fld}, denotes the value of \texttt{obj}'s field called \texttt{fld}.
These are the \emph{accessor} operations. Their FOL representation (as terms) is
$\mathit{index}(a,i)$ and $\mathit{fld}(\mathit{obj})$, respectively.

This mapping allows to formulate predicates on JSON data in FOL.
For example, the guard \verb+db.status.paid <> true+ in \secref{sec:example}'s example maps to the formula $\mathit{paid}(\mathit{status}(\db)) \neq true$.
We also support \emph{updator} operations for both arrays and objects.
For arrays, we have $\mathit{update}(a,i,v)$, which denotes an array that is everywhere the same as $a$ except that at index $i$ it has value $v$.
For objects, we have analogous updator functions per field.
If an object type had fields \texttt{fld1}, \texttt{fld2} \etc, we would then have the term $\mathit{upd\_fld1}(\mathit{obj},v)$, denoting an object everywhere the same as $\mathit{obj}$ except with value $v$ for its field \texttt{fld1}.
We note that these mappings can be automated without effort.
With field and array updators to hand, we can translate a model's scripts (Groovy fragments on graph-edges) into a logical form.
This translation is to a term of one free variable $\db$, denoting the effect of that script on $\db$.

Because standard FOL theorem provers do not natively support the theory of arrays and
objects, we generate suitable FOL axioms from the given JSON type model. %
For arrays, the appropriate axioms are well-known and for objects, there are analogous axioms.
For example,
$\mathit{fld1}(\mathit{upd\_fld1}(\mathit{obj},v)) = v$, and
$\mathit{fld2}(\mathit{upd\_fld1}(\mathit{obj},v)) = \mathit{fld2}(\mathit{obj})$.

In addition, we have concrete syntax for writing complete values (\eg, \texttt{[2,4,6]} for a list of three elements), though this is actually just syntactic sugar for a chain of updates over some underlying base object.
In particular, any database has a (FOL) term representation, called ``database as a term'' below.
Moreover, this same term language allows us to give \emph{partial} specifications of filled databases.
For example, the term $\mathit{upd\_gold}(\sfdb, true)$ stands for a (any) database represented by the constant $\sfdb$ whose \texttt{gold} field holds the value \texttt{true}, with the other fields arbitrary.
Indeed, analysing such partially filled databases is one of the main goals of our research agenda.

\section{Modelling Processes}
\label{sec:process}
In this section we describe our framework for modelling processes. As said earlier,
it is centered around the notion of \emph{process fragments} that manipulate
databases over time. The cooperation of the fragments is described by
\emph{(temporal) constraints}. All constraints and guards in state transitions may refer to user-specified
predicates on (components of) the database, which we call \emph{(logical)
  definitions} here. We will introduce these components now.

\subsection{Process Fragments}
A \emph{guard} $\mu$ is a FOL formula with free variables at most $\{\db\}$;
an \emph{update term} $u$ is a FOL term with free variables at most $\{\db\}$. By
$\Guard$ ($\Update$) we denote the set of all guards (update terms); $\GProg$ is the
set of all Groovy programs. Without further formalization we assume the
Groovy programs are ``sensible'' and describe database updates that can be
characterized as update terms.

A \emph{process fragment} $F$ is directed labeled graph $(N,E,\lambda^\mathsf{N}, \lambda^\mathsf{E})$, where
$N$ is a set of \emph{nodes}, $E \subseteq N \times N$ is a set of \emph{edges},  $\lambda^\mathsf{E}: E \mapsto \Guard
\times \GProg \times \Update$ is an \emph{edge labeling function}, and
$\lambda^\mathsf{N}: N \mapsto 2^{\{ \init, \entry, \exit \} \cup \Guard}$  is an \emph{edge labeling function}.

The informal semantics of process fragments has been given in
Section~\ref{sec:example} already.  The precise semantics of a set of process
fragments is given by first translating it into one single \emph{process model} $\calP$
and then defining the semantics of $\calP$ in terms of its runs.

More formally, a \emph{process (model)} $\calP$ is a quadruple $(N, n_0, E, \lambda^\mathsf{E})$
where $N$, $E$ and $\lambda^\mathsf{E}$ are as above and $n_0 \in N$ is the \emph{initial node}.
Suppose as given a set $\calF = \{ F_1,\ldots,F_k\}$ of process fragments, for some $k \ge
1$, where $F_i = (N_i,E_i,\lambda^\mathsf{N}_i, \lambda^\mathsf{E}_i)$ and $N_i$ and $N_j$
are disjoint, for all $i\neq j$. Suppose further, without loss of generality, that exactly one node in
$\bigcup_{1\le i \le k} N_i$ is labeled as an \init node. Let $n_0$ be that node.
The \emph{process model $\calP = (s_0, n_0, N,  E, \lambda^\mathsf{E})$ associated to $\calF$} is defined as follows:
\vspace*{-1ex}
\begin{align*}
  N &= \textstyle \bigcup_{1\le i \le k} N_i  &   E &= (\textstyle \bigcup_{1\le i \le k} E_i) \cup E^+ &
  \lambda^\mathsf{E} &= (\textstyle \bigcup_{1\le i \le k} \lambda^\mathsf{E}_i) \cup  \lambda^+
\end{align*}
where ($\epsilon$ denotes the empty Groovy program)
\begin{align*}
  E^+ &= \{ (m,n) \mid
   m \in  N_i \text{, } n \in  N_j \text {, }
  \exit \in  \lambda^\mathsf{N_i}(m) \text { and }
  \entry \in   \lambda^\mathsf{N_j}(n) \text{, for some } 1 \le i,j \le k \}\\[-0.3ex]
 \lambda^+ &= \{ (m,n) \mapsto (\gamma,\epsilon,\db) \mid (m,n) \in E^+\text{ and }
               \{\entry, \gamma\}  \subseteq  \lambda^\mathsf{N_j}(n) \text{, for some } 1 \le j \le k
               \}
\end{align*}
For the above construction to be well-defined we require that every \entry node in
every fragment $F_i$ is also labeled with a guard $\gamma$ (which could be $\top$).

\subsection{Definitions and Constraints}
Definitions are logical abbreviations. As such, they are not semantically necessary.
Nonetheless, just as in mathematics, they are a crucial aid in the construction and
comprehensibility of useful models. Formally,
a \emph{definition (for $p$)} is a closed formula of the form $\forall  x\text{:}s\ .\ p(x) \Leftrightarrow \phi[x]$ where $x$
is list of variables of sorts $s \subseteq S$, $p$ is a predicate symbol of the
proper arity, and $\phi$ is a formula.

Constraints specify how process fragments can be combined. The idea has been pursued
before, \eg, in the Declare system~\cite{DBLP:conf/bpm/PesicA06} which uses \emph{propositional}
(linear) temporal logic for that. In order to take data into account, we work with a
fragment of \CTLs over first-logic, which we refer to as \CTLsFO.
The syntax of our \CTLsFO state formulae is given by
$
\Phi ::= \zeta \mid \neg\Phi \mid \Phi \wedge \Phi \mid \pqA\psi \mid \pqE\psi,
$
where $\zeta$ is a FO formula with free variables at most $\{db\}$, and $\psi$ a path formula defined via
$
\psi ::= \Phi \mid \neg \psi \mid \psi \wedge \psi \mid \ltlX \psi \mid \ltlWX \psi \mid \psi \ltlU \psi
$. (The operator $\ltlWX$ is ``weak next''.)
A \emph{constraint} then is simply a state formula. Notice that because constraints may
contain the free variable $\db$, our logic is \emph{not} obtained from propositional
\CTLs by replacing propositional variables by closed formulas.

Figure~\ref{fig:purchase} contains some examples of definitions and constraints.

\subsection{Specifications and Semantics}
The modelling components describing so far are combined into \emph{specifications}.
Formally, a \emph{specification} $\calS$ is a tuple $(\calP,
\calD, \calC)$ where $\calP$ is a process, $\calD$ is a set of definitions and
$\calC$ is a set of constraints.
An \emph{instance $\calI$ (of $\calS$)} is a pair $ (s_0,\calS)$, where $s_0$
is a database (as a term) and $\calS$ is a specification.

We are now in the position to provide a formal definition for the model checking
problems stated in the introduction.
Let $\calS = (\calP, \calD, \calC)$ be as above, where $\calP$
is of the form $(N, n_0, E, \lambda^\mathsf{E})$ and $\phi$ a state formula with free variables
at most $\{\db\}$, the \emph{query}.

As a first step to define the satisfaction relation $(s_0,\calS) \models \phi$ between an
instance and a query we make
the constraints $\calC$ part of the query. Assume $\phi$ is given in negation
normal form (this is always possible) and that it starts with a path quantifier
($\pqE$ or $\pqA$). The \emph{expanded query $\phi_\calC $} is the formula
$\pqA\, (\calC \Rightarrow  \psi)$ if $\phi = \pqA\, \psi$, for some formula $\psi$, and it is
$\pqE\, (\calC \wedge \psi)$ if $\phi = \pqE\, \psi$.
Here, $\calC$ is read as a conjunction of its elements. (The rationale for this definition is that the
desired treatments of constraints is indicated by the path quantifier in the query.)
Notice that with $\phi$ also $\phi_\calC$ is a query.
Now define $(s_0,\calS) \models \phi$ iff $(s_0,\calP, \calD) \models \phi_\calC$, \ie,
the triple $(s_0,\calP, \calD)$ \emph{satisfies} $\phi_\calC$.
It remains to define the latter satisfaction relation, which we turn to now.

As a convenience, we say that \emph{$\calP$ contains a transition $m \stackrel{\gamma, u}
  \longrightarrow n$}.  if $(m,n) \in E$ and $\lambda^\mathsf{E}(m,n) = (\gamma,u)$, for some
guard $\gamma$ and Groovy program $u$ as an update term.

A \emph{run $r$ (of $(\calP, \calD)$) from $s_0$} is a possibly
infinite sequence $(n_0,s_0) (n_1,s_1) (n_2,s_2) \cdots$ of pairs of nodes and
databases, also called \emph{states}, such that (i)
$\calP$ contains transitions of the form $(n_i \stackrel{\gamma_i, u_i} \longrightarrow
n_{i+1})$ , (ii) ${}\models \calD \Rightarrow \gamma_i[s_i]$ and (iii) $s_{i+1} = u_i[s_i]$. In item (i)
in case $i = 0$ the node $n_0$ is meant to be the initial node $n_0$ in $\calP$.
Notice that in item (ii) the definitions $\calD$ play the role of axioms from which
the instantiated guard $\gamma_i[s_i]$ is to follow.
Occasionally the nodes in a run are not important. and we confuse a run with its
projection on the states $s_0 s_1 s_2\cdots$ .

For a run $r = (n_0,s_0)(n_1,s_1)(n_2,s_2) \cdots$  and $i \ge 0$ we define $r[i] = (n_i,s_i)$,
sometimes also $r[i] = s_i$.
By $r^i$ we denote the truncated run $(r_i, s_i)(r_{i+1},
s_{i+1})\cdots$, by $|r|$ the number of elements in the run or $\infty$,
if $r$ is, in fact, infinite.  Obviously, $r^0 = r$.

For any formula $\phi \in \CTLsFO$ with free variables at most $\{\db\}$ we define
$(s_0, \calP, \calD) \models \phi$ as follows:
\[
\begin{array}{lcl}
  (s_0, \calP, \calD) \models \zeta                & \hbox{iff} & {}\models (\calD \Rightarrow \zeta[s_0])\\
  (s_0, \calP, \calD) \models \neg\psi             & \hbox{iff} & (s_0, \calP, \calD) \models \psi \hbox{ is not true}\\
  (s_0, \calP, \calD) \models \psi_1 \wedge \psi_2 & \hbox{iff} & (s_0, \calP, \calD) \models \psi_1 \hbox{ and } (s_0, \calP, \calD) \models \psi_2\\
  (s_0, \calP, \calD) \models \pqA\psi             & \hbox{iff} & (\calP, \calD, r) \models \psi \hbox{ for all runs } r \hbox{ starting in } n_0\\
  (s_0, \calP, \calD) \models \pqE\psi             & \hbox{iff} & (\calP, \calD, r) \models \psi \hbox{ for some run } r \hbox{ starting in } n_0,
\end{array}
\]
where the relation $(\calP, \calD, r) \models \psi$ is defined as
\[
\begin{array}{lcl}
  (\calP, \calD, r) \models \Phi                  & \hbox{iff} & (s_0, \calP, \calD) \models \Phi \\
  (\calP, \calD, r) \models \neg \psi'            & \hbox{iff} & (\calP, \calD, r) \models \psi' \hbox{ is not true}\\
  (\calP, \calD, r) \models \psi'_1 \wedge \psi'_2 & \hbox{iff} & (\calP, \calD, r) \models \psi'_1 \hbox{ and } (\calP, \calD, r) \models \psi'_2\\
  (\calP, \calD, r) \models \ltlX\psi'            & \hbox{iff} & |r| > 1 \hbox{ and } (\calP, \calD, r^1) \models \psi'\\
  (\calP, \calD, r) \models \ltlWX\psi'           & \hbox{iff} & |r| \leq 1, \hbox{ or } |r| > 1 \hbox{ and } (\calP, \calD, r^1) \models \psi'\\
  (\calP, \calD, r) \models \psi'_1 \ltlU \psi'_2  & \hbox{iff} & \hbox{there exists a $j \geq 0$, such that } |r| > j \hbox{ and } (\calP, \calD, r^j) \models \psi'_2,\\
                                           &            &      \hbox{ and } (\calP, \calD, r^i) \models \psi'_1 \hbox{ for all } 0 \leq i < j\\
  (\calP, \calD, r) \models \psi'_1 \ltlR \psi'_2  & \hbox{iff} &  (\calP, \calD, r^i) \models \psi'_2 \hbox{ for all }i \leq |r|, \hbox{ or there exists a } j \geq 0, \hbox{ such that } \\
                                           &             & |r| > j, (\calP, \calD, r^j) \models \psi'_1 \hbox{ and } (\calP, \calD, r^i) \models \psi'_1 \hbox{ for all } 0 \leq i \leq j.
\end{array}
\]
We further assume the usual ``syntactic sugar'', such as $\vee$,
$\Rightarrow$ (implies), $\ltlG$ (always), $\ltlF$ (eventually), or
$\ltlWU$ (weak until) operators, which can easily be defined in terms of
the above set of operators in the expected way.
Note that we distinguish a strong next operator, $\ltlX$, from a weak
next operator, $\ltlWX$ as described in
\cite{bauer:leucker:schallhart:jlc10}.  This gives rise to the following
equivalences: $\psi \ltlR \Phi = \Phi \wedge (\psi \vee \ltlWX \psi
\ltlR \Phi)$ and $\psi \ltlU \Phi = \Phi \vee \psi \wedge \ltlX\psi
\ltlU \Phi$ as one can easily verify by using the above semantics.  This
choice is motivated by our bounded model checking algorithm, which has
to evaluate \CTLsFO formulae over finite traces as opposed to infinite
ones.  For example, when evaluating a safety formula, such as
$\ltlG\psi$, we want a trace of length $n$ that satisfies $\psi$ in all
positions $i \leq n$ to be a model of said formula.  On the other hand,
if there is no position $i \leq n$, such that $\psi'$ is satisfied, we
don't want this trace to be a model for $\ltlF\psi'$.  This is achieved
in our logic as $\ltlG\psi = \psi \wedge \ltlWX\ltlG\psi$ and $\ltlF\psi
= \psi \vee \ltlX\ltlF\psi$ hold.  Note also that $\neg\ltlX\psi \neq
\ltlX\neg\psi$, but $\neg\ltlX \psi = \ltlWX\neg\psi$.

\section{Reasoning with Tableaux for \CTLsFO}
\label{sec:ctlsfo-tableaux}
Tableau calculi for temporal logics have been considered for a long
time~\cite[e.g.]{gore-tableau-methods} as an appropriate and natural reasoning
procedure. There is also a version for propositional
\CTLs~\cite{DBLP:conf/fm/Reynolds09}. However, we are not aware of a first-order
logic tableaux calculus that accommodates our requirements, hence we devise one, see
below. We note that we circumvent the difficult problem of loop detection by working
in a \emph{bounded} model checking setting, where runs are artificially terminated
when they become too long.

Suppose we want to solve an unrestricted model checking problem, \ie, to show that
$(s_0,\calP, \calD) \models \phi_\calC$ holds, for every database $s_0$. As usual with tableau
calculi, this is done by attempting to construct a countermodel for the negation of this
statement. The universally quantified database $s_0$ then becomes a Skolem
constant, say, $\sfdb$, representing an (unknown) initial database.
A \emph{state} then is a pair of the form $(n,u[\sfdb])$ where $n
\in N$ and $u[\sfdb]$ is an update term instantiated with that initial database. We
find it convenient to formulate the calculus' inference rules as operators on (sets
of) sequents. A \emph{sequent} is an expression of the form
$s \vdash_Q \Phi$ where $s$ is a state, $Q \in \{\pqE, \pqA\}$ is a path quantifier, and $\Phi[\db]$ is
a (possible empty) set of \CTLsFO formulas in negation normal form with free variables at most
$\{ \db \}$. When we write $s \vdash_Q \phi,\Phi$ we mean $s \vdash_Q \{\phi\} \cup\Phi$.

The informal semantics of a sequent $(n,u[\sfdb]) \vdash_Q \Phi[\db]$ is
``some run of the instance $ (\sfdb, \calP, \calD)$ has reached the state $(n,u[\sfdb])$ and
$(n,u[\sfdb]) \models Q\, \Phi[u[\sfdb]]$''.

A \emph{tableau} calculus, the calculus below derives trees that represent
disjunctions of conjunctions of formulas. More precisely, the nodes are labeled with
sets of sequents that are read conjunctively, and sibling nodes are connected
disjunctively.  The purpose of the calculus' inference rules is to analyse a given
sequent by breaking up the formulas in the sequent according to their boolean
operators, path quantifiers and temporal operators. An additional implicit and/or
structure is given by reading the formulas $\Phi$ in $s \vdash_\pqE \Phi$ conjunctively, and
reading the formulas $\Phi$ in $s \vdash_\pqA \Phi$ disjunctively. The reason is that $\pqA$
does not distribute over ``or'' and $\pqE$ does not distribute over ``and''.

We need some more definitions to formulate the calculus.
A formula is \emph{classical} iff it contains no path quantifer and no temporal
operator. A formula is a \emph{modal atom} iff its top-level operator is a path
quantifer or a temporal operator.
A sequent $s \vdash_Q \Phi$  is \emph{classical} if all formulas in $\Phi$  are classical.

A \emph{tableau node} is a (possibly empty) set of sequents, denoted by the letter
$\Sigma$.  We often write $\sigma ;\Sigma$  instead of $\{\sigma \} \cup  \Sigma$. We simply speak of ``nodes''
instead of ``tableau nodes'' if confusion with the nodes in graphs
is unlikely.

Let $\phi_\calC$  be a given expended query and $\calS$ a specification as introduced
before. The \emph{initial sequent} is the
sequent $s_0 \vdash_\pqE \neg \phi_\calC$, where $s_0 = (n_0, \sfdb)$ is the \emph{initial state}, for some
fresh constant $\sfdb$. Notice that the expanded query is negated, corresponding to the
intuition of attempting to compute a countermodel for the negation of the expanded
query.

Because we are adopting a standard notion of tableau derivations it suffices to
define the inference rules. (The root node contains the initial sequent only.)
The components $\calP$ and $\calD$ are left implicit below.

\paragraph{Boolean rules.}
The implicit reading of $\Phi$  as disjunctions/conjunctions in a
$\vdash_\pqA$/$\vdash_\pqE$ sequent sanction the following rules.

\begin{displaymath}
\begin{matrix}
\infrule[$\pqE$-$\land$]{
	   s \vdash_\pqE \phi  \land  \psi , \Phi  ; \Sigma
}{ 
	   s \vdash_\pqE \phi , \psi , \Phi  ; \Sigma
}
& \qquad
\infrule[$\pqE$-$\lor$]{	           s \vdash_\pqE \phi  \lor  \psi , \Phi  ; \Sigma
}{
	   s \vdash_\pqE \phi , \Phi  ; \Sigma\quad        s \vdash_\pqE \psi , \Phi  ; \Sigma
} \\[3ex]
\infrule[$\pqA$-$\lor$]{
	   s \vdash_\pqA \phi  \lor  \psi , \Phi  ; \Sigma
}{
	   s \vdash_\pqA \phi , \psi , \Phi  ; \Sigma
}
& \qquad
\infrule[$\pqA$-$\land$]{	   s \vdash_\pqA \phi  \land  \psi , \Phi  ; \Sigma
}{
	   s \vdash_\pqA \phi,\Phi ;  s \vdash_\pqA \psi , \Phi  ; \Sigma
}
\end{matrix}
\end{displaymath}
if $\phi$  is not classical or $\psi$  is not classical (no need to
break classical formulas apart).

\paragraph{Rules to separate classical sequents.}
The following rules separate away the classical formulas from the modal atoms in $\Phi$.
Every classical sequent can be passed on to a
first-order theorem prover; if the result is ``unsatisfiable'' then the node is closed.
\begin{displaymath}
  \begin{matrix}
\infrule[$\pqE$-Split]{	          s \vdash_\pqE \Phi  ; \Sigma
}{
	          s \vdash_\pqE \Gamma[u[\sfdb]]  ;  s \vdash_\pqE \Phi \backslash \Gamma  ; \Sigma
}
& \qquad
\infrule[$\pqA$-Split]{	                s \vdash_\pqA \Phi  ; \Sigma
}{
	    s \vdash_\pqA \Gamma[u[\sfdb]]  ; \Sigma  \qquad	    s \vdash_\pqA \Phi \backslash \Gamma  ; \Sigma
}
  \end{matrix}
\end{displaymath}
if $s = (n,u[\sfdb])$ for some $n$,  $\Gamma$  consists of all classical formulas in $\Phi$,
$\Gamma[u[\sfdb]]$ is obtained from $\Gamma$ by replacing every free occurence of the
variable $\db$ in all its formulas by $u[\sfdb]$, and $\Gamma \neq \emptyset$ and $\Gamma[u[\sfdb]] \neq \Phi$.

The left rule exploits the equivalence $\pqE(\phi  \land  \Phi ) \equiv  \pqE \phi  \land  \pqE \Phi$  if $\phi$
is classical, and the right rule exploits the equivalence
$\pqA(\phi  \lor  \Phi ) \equiv  \pqA \phi  \lor  \pqA \Phi$  if $\phi$  is classical.

\paragraph{Rules for path quantifiers.}
—————————————————————————–
The next rules eliminate path quantifiers, where $Q \in  \{\pqE, \pqA\}$.

\begin{displaymath}
\begin{matrix}
\infrule[$\pqE$-Elim]{                s \vdash_\pqE Q\,\phi , \Phi  ; \Sigma
}{
                s \vdash_Q \phi  ; s \vdash_\pqE \Phi  ; \Sigma
}
& \qquad
\infrule[$\pqA$-Elim]{	                      s \vdash_\pqA Q\,\phi , \Phi  ; \Sigma
}{
                s \vdash_Q \phi  ; \Sigma\qquad             s \vdash_\pqA \Phi  ; \Sigma
}
\end{matrix}
\end{displaymath}
The soundness of the left rule follows from the equivalences $\pqE\,(Q\,\phi  \land  \Phi ) \equiv
\pqE\, Q\,\phi  \land  \pqE\,\Phi  \equiv  Q\,\phi  \land  \pqE\,\Phi$, and
the soundness of the right rule follows from the equivalences
$\pqA\,(Q\,\phi  \lor  \Phi ) \equiv  \pqA\, Q\,\phi  \lor  \pqA\,\Phi  \equiv  Q\,\phi  \lor  \pqA\,\Phi$.

The above rules apply also if $\Phi$ is empty. Notice that in this case $\Phi$ represents
the empty conjunction in $s \vdash_\pqE\, \Phi$, which is equivalent to $\top$, and the empty
disjunction in $s \vdash_\pqA\, \Phi$, which is equivalent to $\bot$.

When applied exhaustively, the rules so far lead to sequents that all have the form
$s \vdash_Q \Phi$  such that (a) $\Phi$  consists of classical formulas only, or
(b) $\Phi$  consists of modal atoms only with top-level operators from $\{ \toU, \toR, \toX, \toWX \}$.

\paragraph{Rules to expand $\toU$ and $\toR$ formulas.}
The following rules perform one-step expansions of modal atoms with $\toU$ and $\toR$ operators.
\begin{displaymath}
\begin{matrix}\small
\small\infrule[$\toU$-Exp]{	      s \vdash_Q (\phi\,  \toU\, \psi ), \Phi  ; \Sigma
}{
	      s \vdash_Q \psi  \lor  (\phi  \land  \toX\,(\phi\,  \toU\, \psi )), \Phi  ; \Sigma
}
&\qquad
\small\infrule[$\toR$-Exp]{	      s \vdash_Q (\phi\,  \toR\, \psi ), \Phi  ; \Sigma
}{
	      s \vdash_Q (\psi  \land  (\phi  \lor  \toWX\,(\phi\,  \toR\, \psi ))), \Phi  ; \Sigma
}
\end{matrix}
\end{displaymath}
When applied exhaustively, the rules so far lead to sequents that all have the form
$s \vdash_Q \Phi$  such that (a) $\Phi$  consists of classical formulas only, or
$\Phi$  consists of modal atoms only with top-level operators from $\{ \toX, \toWX \}$.

\paragraph{Rules to simplify $\toX$ and $\toWX$ formulas.}
Below we define inference rules for one-step expansions of sequents of the form
$s \vdash_Q \toX\, \phi$ and $\vdash_Q \toWX\, \phi$. The following inference rules prepare their
application.
\begin{displaymath}
\begin{matrix}
\infrule[$\pqE$-$\toX$-Simp]{	      s \vdash_\pqE \toX\,\phi _1 , \ldots , \toX\,\phi _n, \toWX\,\psi _1 , \ldots , \toWX\,\psi _m ; \Sigma
}{
	      s \vdash_\pqE Y\,(\phi _1  \land  \cdots  \land  \phi _n \land  \psi _1  \land  \cdots  \land  \psi _m) ; \Sigma
}
\end{matrix}
\end{displaymath}
if $n+m>0$, where $Y = \toWX$ if $n = 0$ else $Y = \toX$.
Intuitively, if just one of the modal atoms in the premise is an $\toX$-formula then
a successor state must exist to satisfy it, hence the $\toX$-formula in the
conclusion.
Similarly:
\begin{displaymath}
\begin{matrix}
\infrule[$\pqA$-$\toX$-Simp]{ s \vdash_\pqA \toX\,\phi _1 , \ldots , \toX\,\phi _n, \toWX\,\psi _1 , \ldots , \toWX\,\psi _m ; \Sigma
}{
	      s \vdash_\pqA Y(\phi _1  \lor  \cdots  \lor  \phi _n \lor  \psi _1  \lor  \cdots  \lor  \psi _m) ; \Sigma
}\end{matrix}
\end{displaymath}
if $n+m>0$, where $Y = \toX$ if $m = 0$ else $Y = \toWX$.

The correctness of this rule follows from the equivalences
$\pqA\,(\toX\,\phi  \lor \,\toWX\,\psi ) \equiv \pqA\,(\toWX\,\phi  \lor  \toWX\,\psi ) \equiv \pqA\,\toWX\,(\phi  \lor  \psi )$.

To summarize, with the rules so far, all sequents can be brought into one
of the following forms:
(a) $s \vdash_Q \Gamma$, where $\Gamma$  consists of classical formulas only,
(b) $s \vdash_Q \toX\,\phi$, or
(c) $s \vdash_Q \toWX\,\phi$.

\paragraph{Rule to close branches.}
The following rule derives no conclusions and this way indicates that a branch in a
tableau is ``closed''.
\begin{displaymath}
\begin{matrix}
\infrule[Unsat]{    s_1  \vdash_{Q_1}  \Phi_1  ; \cdots  ; s_n \vdash_{Q_n} \Phi_n
}{
               \mbox{}
}
\end{matrix}
\end{displaymath}
if every $\Phi_i$ consists of closed classical formulas, and
$\bigwedge (\calD \cup \Phi _1 \cup \cdots \cup \Phi _n)$ is unsatisfiable (not satisfiable).

\paragraph{Rules to expand $\toX$ and $\toWX$ formulas.}
\begin{displaymath}
\begin{matrix}
\small\infrule[$\pqE$-$\toWX$-Exp]{
       (m,t) \vdash_\pqE \toWX\,\phi  ; \Sigma
}{
      (n_1, u_1[t]) \vdash_\pqE \gamma_1[t] \land \phi  ; \Sigma  \quad \cdots \quad
      (n_k, u_k[t])  \vdash_\pqE \gamma_k[t] \land  \phi ; \Sigma \quad
      (m,t) \vdash_\pqE \lnot \gamma_1[t] \land  \cdots  \land  \lnot \gamma_k[t] ; \Sigma
}
\end{matrix}
\end{displaymath}
if there is a $k \ge 0$ such that $m \stackrel{\gamma_i,u_i}{\longrightarrow} n_i$ are all
transitions in $\calP$ emerging from $m$, where $1 \le i \le k$.

This rule binds the variable $\db$ in the guards to the term $t$,which represents the
current database, while it leaves the formula $\phi$ untouched.  The variable
$\db$ in $\toWX\, \Phi$ refers to the databases in the successor states, \ie, the
databases $u_i[t]$. The rules to separate classical sequents above will
bind $\db$ in $\Phi$ correctly.

There is also a rule \IR{$\pqE$-$\toX$-Exp} whose premise sequent is made with the
$\toX$ operator instead of $\toWX$. It differs from the \IR{$\pqE$-$\toWX$-Exp} rule
only by leaving away the rightmost conclusion. We do not display it here for space reasons.
We note that both rules are defined also if $k=0$.

\begin{displaymath}
\begin{matrix}
\infrule[$\pqA$-$\toX$-Exp]{
       (m,t) \vdash_\pqA \toX\,\phi  ; \Sigma
}{
      (n_1, u_1[t]) \vdash_\pqA \lnot \gamma_1[t] \vee \phi ;
      \cdots
      (n_k, u_k[t]) \vdash_\pqA \lnot \gamma_k[t] \vee \phi ;
      (m,t) \vdash_\pqE \gamma_1[t] \vee  \cdots  \vee \gamma_k[t] ; \Sigma

}
\end{matrix}
\end{displaymath}
if there is a $k \ge 0$ such that $m \stackrel{\gamma_i,u_i}{\longrightarrow} n_i$ are all
transitions in $\calP$ emerging from $m$, where $1 \le i \le k$.

This rule will for each of the conclusion sequent lead to a case
distinction (via branching) whether the guard of a transition is true
or not. Only if the guard is true the transition must be taken.
The conclusion sequent $(m,t) \vdash_\pqE \gamma_1[t] \vee  \cdots  \vee \gamma_k[t]$ forces that at least one guard is
true. Analogously to above, there is also a rule \IR{$\pqA$-$\toWX$-Exp} for the
$\toWX$ case, which does not include this sequent. This reflects that 
$\toWX$ formulas are true in states without successor.

Both rules also work as expected if $k=0$: for \IR{$\pqA$-$\toX$-Exp} 
the formula in the sequent $(m,t) \vdash_\pqE \gamma_1[t] \vee  \cdots  \vee \gamma_k[t]$ is equivalent to $\bot$
(false); for \IR{$\pqA$-$\toWX$-Exp} the premise sequent is 
deleted. If additionally $\Sigma$ is empty then the result is a node with the empty set
of sequents. This does not indicate branch closure; branch closure is
indicated by deriving \emph{no} conclusions, not a unit-conclusion, even if empty.

This concludes the presentation of the tableau calculus. 
As said above, we enforce
derivations to be finite by imposing a user-specified maximal length on the number of
state transitions it executes. This is realized as a check
in the  rules to expand $\toX$ and $\toWX$ formulas by pretending a value $k=0$ of
transitions emerging from the node of the considered state, if the run to that state
becomes too long. (This is not formalized above.) 

For this  bounded model checking setting we obtain a formal soundness and completeness result for
the (hence, bounded) unrestricted model checking problem. More precisely, 
given a specification $\calS = (\calP, \calD, \calC)$, 
$(s_0,\calS) \models \Phi$ holds for every database $s_0$  relative to all runs of
maximal length shorter than a given finite length $l$ if and only if the fully expanded tableau with 
initial node $(n_0, \sfdb) \vdash_\pqE \phi_\calC$ is closed. (A tableau is closed if each of
its leafs is closed as determined by the \IR{Unsat} rule or the \IR{$E$-$\toX$-Exp} rule.)

The \IR{Unsat} tableau rule requires a call to a (sound) first-order theorem
prover. Depending on the underlying syntactic fragment of FOL these calls may not
always terminate. However, if a classical sequent is provably \emph{satisfiable} then
it is possible to extract from the tableaux branch a run that constitutes a
counterexample to the given problem. Moreover, this formula will often represent
\emph{general} conditions on the initial database $s_0$ under which the query $\Phi$ is not satisfied by
$(s_0,\calS)$ and this way provide more valuable feedback than a fully concrete database.

\section{Practice and Experiments}
\label{sec:experiments}

In this section, we provide some notes on the implementations of the theory presented
in the preceding sections.  

\paragraph{Satisfiability Checking on the Nodes.}
Before we can model-check the truth of formulas over the graph structure of a full
specification, we must be able to evaluate first-order formulas with respect to nodes
within that graph.  When performing checking with a concrete initial state, all
subsequent states will be concrete as well, and evaluating quantified formulas is
straightforward as long as quantification is over finite domains, as is typical.
On the other hand, if the initial state is only characterised with a formula, then
checking satisfiability of formulas with respect to that node and all its successors
becomes a full-blown theorem-proving problem.

We solve this problem by translating to the standard TPTP format~\cite{tptp2009},
which has recently be extended to include
arithmetic~\cite{Sutcliffe:etal:TPTP-TFFA:LPAR:2012}, and then using off-the-shelf
first-order provers.  Our current backend is SPASS+T~\cite{SPASST2006}, which has
good support for arithmetic in addition to sorted first-order logic.

\paragraph{Model Checking.}
For concrete model checking, we assume that there are no two definitions for same predicate symbol, that
definitions are not recursive, and that all quantifications inside the bodies $\phi$
range over concrete data items.  With these assumptions, definitions can be expanded
as necessary, and we can efficiently decide if formulas (edges' guards and the
classical sub-formulas of the model checking problem) are satisfied with respect to
concrete database values.  In theory, SPASS+T should do the same, but we have found
that our own custom guard evaluator performs better, and is also guaranteed to
terminate.  When performing concrete model checking, we can also execute scripts
directly as Groovy programs rather than needing to manipulate them as first order
terms.  

We have fully implemented the preceding section's generic tableau system for concrete
model checking, giving us an efficient procedure that is guaranteed to terminate on
problems given a depth-bound. In our practical experiments on the example in
Section~\ref{sec:example} we could (dis)prove queries like the ones mentioned there
in very short time.

Our implementation is also capable of generating proof obligations in the TPTP
format for unbounded model checking. It also emits the necessary axioms to reflect
the semantics of objects and arrays, as explained in Section~\ref{sec:data}. We have
experimented with smaller examples and found that SPASS+T is capable of handling
them. At the current stage, however, the implementation is not mature enough yet, and
so our experiments are too premature to report on. We also plan to consider
alternatives to SPASS+T by implementing the calculus
in~\cite{Baumgartner:Tinelli:MEET:CADE:2011} and by linking in SMT-solvers.

\section{Conclusions and Future Work}
\label{sec:conclusion}
We described a novel approach to modelling and reasoning about data-centric business
processes. Our modelling language treats data, process fragments, constraints and
logical definitions of business rules on a par. Our research plan focuses on
providing strong analytical capabilities on the corresponding models by taking all
these components into account. The main ambition is to go beyond model checking from
concrete initial states. To this end we have devised a novel tableau calculus that reduces
what we called unrestricted model checking problems to first-order logic over
arithmetic. 

Our main contributions so far are conceptual in nature. Our main theoretical result
is the soundness and completeness of the tableau calculus, as explained at the end of
Section~\ref{sec:process}.  Our implementation is already fully functional for
concrete model checking.

Much remains to be done, at various levels. The tableau implementation needs to be
completed and improved for efficiency, and more experiments need to be carried out. 

The main motivation for using JSON and Groovy is their widespread acceptance in
practice and available tool support, which we exploit in our implementation. For the
same reason we want to extend our modelling language by front-ends for established
business process modeling techniques, in particular BPMN. This raises (also) some
non-trivial interesting theoretical issues. For example, how to map BPMN's
parallel-And construct into our framework. We expect that by using process \emph{fragments}
and constraints on them an isomorphic mapping is possible.


\bibliographystyle{abbrv}
\bibliography{bibliography}

\begin{thebibliography}{10}

\bibitem{bauer:leucker:schallhart:jlc10}
A.~Bauer, M.~Leucker, and C.~Schallhart.
\newblock Comparing {LTL} semantics for runtime verification.
\newblock {\em Logic and Computation}, 20(3):651--674, 2010.

\bibitem{Baumgartner:Tinelli:MEET:CADE:2011}
P.~Baumgartner and C.~Tinelli.
\newblock Model evolution with equality modulo built-in theories.
\newblock In N.~Bjoerner and V.~Sofronie-Stokkermans, editors, {\em CADE-23 --
  The 23nd International Conference on Automated Deduction}, volume 6803 of
  {\em Lecture Notes in Artificial Intelligence}, pages 85--100. Springer,
  2011.

\bibitem{clarke_em-etal:1999a}
E.~M. Clarke, O.~Grumberg, and D.~A. Peled.
\newblock {\em Model Checking}.
\newblock The {MIT} Press, Cambridge, Massachusetts, 1999.

\bibitem{JSON}
D.~Crockford.
\newblock {RFC 4627}---{T}he application/json media type for {JavaScript}
  {Object} {Notation} ({JSON}).
\newblock Technical report, IETF, 2006.

\bibitem{DBLP:conf/bpm/DamaggioDHV11}
E.~Damaggio, A.~Deutsch, R.~Hull, and V.~Vianu.
\newblock Automatic verification of data-centric business processes.
\newblock In S.~Rinderle-Ma, F.~Toumani, and K.~Wolf, editors, {\em BPM},
  volume 6896 of {\em Lecture Notes in Computer Science}, pages 3--16.
  Springer, 2011.

\bibitem{gore-tableau-methods}
R.~Gor\'{e}.
\newblock Chapter~6: Tableau methods for modal and temporal logics.
\newblock In {M {D'Agostino}, D Gabbay, R H\"{a}hnle, J Posegga}, editor, {\em
  Handbook of Tableau Methods}, pages 297--396. Kluwer Academic Publishers,
  1999.

\bibitem{NYT120212}
S.~Lohr.
\newblock {The Age of Big Data}.
\newblock {Sunday Review}, New York Times, Feb. 2012.

\bibitem{DBLP:journals/ibmsj/NigamC03}
A.~Nigam and N.~S. Caswell.
\newblock Business artifacts: An approach to operational specification.
\newblock {\em IBM Systems Journal}, 42(3):428--445, 2003.

\bibitem{DBLP:conf/bpm/PesicA06}
M.~Pesic and W.~M.~P. van~der Aalst.
\newblock A declarative approach for flexible business processes management.
\newblock In J.~Eder and S.~Dustdar, editors, {\em Business Process Management
  Workshops}, volume 4103 of {\em Lecture Notes in Computer Science}, pages
  169--180. Springer, 2006.

\bibitem{SPASST2006}
V.~Prevosto and U.~Waldmann.
\newblock {SPASS+T}.
\newblock In G.~Sutcliffe, R.~Schmidt, and S.~Schulz, editors, {\em ESCoR:
  Empirically Successful Computerized Reasoning}, volume 192 of {\em CEUR
  Workshop Proceedings}, 2006.

\bibitem{DBLP:conf/fm/Reynolds09}
M.~Reynolds.
\newblock A tableau for {CTL*}.
\newblock In A.~Cavalcanti and D.~Dams, editors, {\em FM}, volume 5850 of {\em
  Lecture Notes in Computer Science}, pages 403--418. Springer, 2009.

\bibitem{Sutcliffe:etal:TPTP-TFFA:LPAR:2012}
G.~S.~S. Schulz, K.~Claessen, and P.~Baumgartner.
\newblock The tptp typed first-order form with arithmetic.
\newblock In N.~Bjoerner and A.~Voronkov, editors, {\em Proceedings of the 18th
  International Conference on Logic for Programming, Artificial Intelligence
  and Reasoning (LPAR-18)}, volume 7180 of {\em Lecture Notes in Artificial
  Intelligence}. Springer, 2012.

\bibitem{tptp2009}
G.~Sutcliffe.
\newblock {The {TPTP} Problem Library and Associated Infrastructure: The {FOF}
  and {CNF} Parts, v3.5.0}.
\newblock {\em Journal of Automated Reasoning}, 43(4):337--362, 2009.

\end{thebibliography}


\end{document}